%% file: main.tex
\documentclass[runningheads]{llncs}

\input{_packages}

\begin{document}

\title{Leveraging AI for Productive and Trustworthy HPC Software: Challenges and Research Directions}

\input{_authors}

\maketitle

\begin{abstract}
We discuss the challenges and propose research directions for using AI to revolutionize the development of high-performance computing (HPC) software.
AI technologies, in particular large language models, have transformed every aspect of software development. For its part, HPC software is recognized as a highly specialized scientific field of its own. We discuss the challenges associated with leveraging state-of-the-art AI technologies to develop such a unique and niche class of software and outline our research directions in the two US Department of Energy--funded projects for advancing HPC Software via AI: Ellora and Durban.

\end{abstract}

\input{01-Introduction}

\input{02-Background}

\input{03-Challenges}

\input{04-Directions}

\input{05-Conclusions}

\section*{Acknowledgments}
This material is based upon work supported by the U.S. Department of Energy, Office of Science, Office of Advanced Scientific Computing Research, through solicitation DE-FOA-0003264, ``Advancements in Artificial Intelligence for Science,'' under Award Numbers DE-SC0025598 and DE-SC0025645.
This manuscript has been authored by UT-Battelle, LLC, under contract DE-AC05-00OR22725 with the US Department of Energy (DOE). The US government retains and the publisher, by accepting the article for publication, acknowledges that the US government retains a nonexclusive, paid-up, irrevocable, worldwide license to publish or reproduce the published form of this manuscript, or allow others to do so, for US government purposes. DOE will provide public access to these results of federally sponsored research in accordance with the DOE Public Access Plan (\url{https://www.energy.gov/doe-public-access-plan}). \,\,
This work performed under the auspices of the U.S. Department of Energy by Lawrence Livermore National Laboratory under Contract DE-AC52-07NA27344 (LLNL-CONF-2005811). \,\,
This manuscript has been authored by an author at Lawrence Berkeley National Laboratory under Contract No. DE-AC02-05CH11231 with the U.S. Department of Energy. The U.S. Government retains, and the publisher, by accepting the article for publication, acknowledges, that the U.S. Government retains a non-exclusive, paid-up, irrevocable, world-wide license to publish or reproduce the published form of this manuscript, or allow others to do so, for U.S. Government purposes.
\bibliographystyle{splncs04}
\bibliography{references}

\end{document}

%% file: _packages.tex
\usepackage[T1]{fontenc}
\usepackage{graphicx}
\usepackage{hyperref}
\usepackage{float}
\usepackage{subcaption}
\usepackage{enumerate}
\usepackage{bbding}

%% file: _authors.tex
\author{
Keita Teranishi\inst{1}
\orcidID{0000-0001-6647-2690} 
\and
Harshitha Menon\inst{2}
\orcidID{0000-0003-4707-9580}
\and
William F. Godoy\inst{1}\Envelope
\orcidID{0000-0002-2590-5178}
\and
Prasanna Balaprakash\inst{1}
\orcidID{0000-0002-0292-5715}
\and
David Bau\inst{5}
\orcidID{0000-0003-1744-6765}
\and
Tal Ben-Nun\inst{2}
\orcidID{0000-0002-3657-6568}
\and
Abhinav Bhatele\inst{6}
\orcidID{0000-0003-3069-3701}
\and
Franz Franchetti\inst{4}
\orcidID{0000-0002-3529-8973}
\and
Michael	Franusich\inst{7}
\orcidID{0000-0002-8190-4292}
\and
Todd Gamblin\inst{2}
\orcidID{0000-0002-7857-2805}
\and
Giorgis Georgakoudis\inst{2}
\orcidID{0000-0001-6542-3555}
\and
Tom Goldstein\inst{6}
\orcidID{0000-0003-1660-9307}
\and
Arjun Guha\inst{5}
\orcidID{0000-0002-7493-3271}
\and
Steven E. Hahn\inst{1}
\orcidID{0000-0002-2018-7904}
\and
Costin Iancu\inst{3}
\orcidID{0000-0001-7845-2427}
\and
Zheming	Jin\inst{1}
\orcidID{0000-0002-7197-780X}
\and
Terry Jones\inst{1}
\orcidID{0000-0003-2187-9707}
\and
Tze-Meng Low\inst{4}
\orcidID{0000-0002-5179-8249}
\and
Het	Mankad\inst{1}
\orcidID{0000-0003-4339-599X}
\and
Narasinga Rao Miniskar\inst{1}
\orcidID{0000-0001-8259-8891}
\and
Mohammad Alaul Haque Monil\inst{1}
\orcidID{0000-0003-3419-4037}
\and
Daniel Nichols\inst{6}
\orcidID{0000-0002-3538-6164}
\and
Konstantinos Parasyris\inst{2}
\orcidID{0000-0002-8258-9693}
\and
Swaroop	Pophale\inst{1}
\orcidID{0000-0001-8544-6367}
\and
Pedro Valero-Lara\inst{1}
\orcidID{0000-0002-1479-4310}
\and
Jeffrey S. Vetter\inst{1}
\orcidID{0000-0002-2449-6720}
\and
Samuel Williams\inst{3}
\orcidID{0000-0002-8327-5717}
\and
Aaron Young\inst{1}
\orcidID{0000-0002-5448-4667}
}

\authorrunning{K. Teranishi , H Menon, W Godoy, et al.}

\titlerunning{Leveraging AI for Prod and Trust HPC SW}

\institute{Oak Ridge National Laboratory, Oak Ridge, Tennessee, USA \\ 
\Envelope \email{godoywf@ornl.gov}
\and
Lawrence Livermore National Laboratory, Livermore, California, USA
\and
Lawrence Berkeley National Laboratory, Berkeley, California, USA
\and
Carnegie Mellon University, Pittsburgh, Pennsylvania, USA
\and
Northeastern University, Boston, Massachusetts, USA
\and
University of Maryland, College Park, Maryland, USA
\and
SpiralGen Inc., Pittsburgh, Pennsylvania, USA
}

%% file: 01-Introduction.tex
\section{Introduction}
\label{sec:Introduction}

As a key government investment, high-performance computing (HPC) has enabled unprecedented advances in science and engineering in recent decades. However, the future of HPC in the post-Moore era faces new challenges, including record-breaking costs in semiconductor manufacturing, an unprecedented number of AI workloads, tighter energy requirements and ever-increasing operational costs~\cite{doi:10.1126/science.adu0801,reed2022reinventinghighperformancecomputing}. 

To fully leverage the capabilities of the latest generation of exascale systems, the US Department of Energy (DOE) made significant investments in a large, interoperable, and performance-portable HPC software ecosystem as part of the Exascale Computing Project
(ECP, 2016--2023)~\cite{kothe2018exascale}. These efforts included the Extreme-Scale Scientific Software Stack (E4S),\footnote{\url{https://e4s.io/}} which is now being supported by the DOE's Advanced Scientific Computing Research (ASCR) program as it continues the development of the exascale software ecosystem. Nevertheless, programming, developing, and deploying robust, trustworthy, and verifiable HPC software has become a complex endeavor, with ever-increasing advancement costs required to meet the science-application needs on new heterogeneous systems and architectures~\cite{osti_1473756}.

AI technologies, in particular state-of-the-art large language models (LLMs), have shown incredible potential in a variety of software tasks~\cite{9849664,chen2021evaluating,10.1145/3524842.3528470,10.1145/3501385.3543957} and can significantly improve programmer productivity~\cite{bird:taking-flight-with-copilot,murali:code-compose,dunay:multiline-code-compose,liang:copilot-survey}. For example, Meta reports that 99\% of their developers use an internal LLM to discover APIs, accelerate their work, and generate 9\%--17\% of all code that they write~\cite{dunay:multiline-code-compose}.
However, advancements in state-of-the-art foundational LLMs, such as the Generative Pre-trained Transformer (GPT-3/4)~\cite{NEURIPS2020_1457c0d6}, Claude 3~\cite{claude3}, and Llama-3~\cite{llama3modelcard}, are driven mainly by industry needs. As a result, despite the popularity of general-purpose LLMs for writing code, our team's pioneering research in this area showed that LLMs are far less effective at HPC programming tasks ~\cite{nicholshpc,nichols2024performancealigned,nichols2024large,3605886,valerolara2023comparing,godoy2024large} because the data used to train the LLMs contains far fewer HPC-specific codes (i.e., there are simply fewer HPC codes vs. general software from which the LLMs can learn)~\cite{joel2024surveyllmbasedcodegeneration,chen2024landscape}. This means that other runtime dimensions (e.g., parallel performance, portability, reproducibility of scientific results) are also not reflected in code/text sources or other AI training corpora.

Integrating AI into HPC software development presents a transformative opportunity to enhance the efficiency, performance, and productivity of the entire software development life cycle. 
Fortunately, DOE's ASCR program has recognized the need to explore the use of AI in advancing the next generation of HPC codes and launched the Advancements in Artificial Intelligence for Science program in 2024 by awarding \$68 million to proposed research projects that leverage AI for scientific innovation. One key area is the ``AI Innovations for Scientific Knowledge Synthesis and Software Development.''
In this paper, we provide an overview of the characteristics of HPC software (Section~\ref{sec:Background}), describe the challenges at the intersection of AI and coding for HPC  (Section~\ref{sec:Challenges}), and propose research directions for an impactful application of AI in the HPC software ecosystem (Section~\ref{sec:Research Directions}). We mainly focus on two important areas in which AI can revolutionize HPC software: productivity and performance portability.

%% file: 02-Background.tex
\section{Background on HPC Software}
\label{sec:Background}

HPC software has been the cornerstone of scientific discovery at scale for decades \cite{dongarra2024co}. This history is built upon a rich legacy of advances in algorithms, math libraries, programming languages, and systems and standards that enable the actual utility provided by the fastest supercomputers in the world. In fact, given their niche nature, HPC codes are closer to scientific instruments than they are to general-purpose software, with their clear goals of advancing the state of the art in computational and computer sciences.

\paragraph{Characteristics:} Below are some of the unique characteristics of HPC software that must be understood before anyone can replicate---for HPC---the impact that AI has had in other software domains:

\begin{itemize}
  \item Massively parallel and scalable up to the largest supercomputers in the world
  \item Vendor neutral/portable to new architectures 
  \item The vast majority is written in C, C\texttt{++}, and Fortran with ecosystem components in high-level languages (e.g., Python, R, Julia, Matlab)
  \item Deep reliance on vendor software stacks such as CUDA and HIP for accelerator platforms
  \item Low developers-to-users ratio (i.e., few developers, many users)
  \item Complex and diverse ecosystem that includes algorithms, programming systems, runtime systems, compilers, tools, applications, libraries, workflows, network, I/O, packaging, and more
  \item Requires knowledge of hardware specifications to balance compute, communication, memory, I/O
  \item Correctness includes runtime characteristics such as performance, portability, scalability, reliability, and scientific accuracy
  \item Driven by impact on scientific discovery, thus not necessarily prioritizing resources for general software use (e.g., documentation, training)
  \item Developed by highly specialized individuals with multidisciplinary scientific backgrounds
  \item Cutting-edge HPC software evolves significantly with every supercomputing architecture (e.g., distributed, many-core, GPU/hybrid)
  \item HPC software has benefited from advancements in open-source technologies, although much of the mission-critical software is still classified or not publicly available 
  \item The vast majority is government funded (e.g., DOE) and does not have for-profit goals
\end{itemize}

\paragraph{Requirements for modern software:}
HPC software relies on the broader software landscape (e.g., programming languages, vendor tools, runtimes). Hence, the HPC stack must evolve accordingly and alongside the broader software ecosystem to meet the post-Moore era's requirements---including trustworthiness (e.g., memory-safety, robustness)~\cite{house2024back}, reproducibility, maintainability, and energy-efficiency---and to align with national interests. Not meeting these requirements puts the field at risk in mission-critical scenarios, and the clearest example of vulnerability is the overwhelming reliance on Fortran~\cite{shipman2023evaluation,9736688} in legacy HPC codes.

%% file: 03-Challenges.tex
\section{AI Challenges for HPC Software}
\label{sec:Challenges}

State-of-the-art AI advancements present numerous opportunities to enhance HPC software. However, bridging HPC and AI requires a concerted effort from the community because it faces several well-documented challenges in scientific computing~\cite{osti_1986455}. This section explores key challenges in leveraging AI across various aspects of HPC software.

\paragraph{Code generation, translation, and optimization:} State-of-the-art foundational LLMs are not trained to generate, translate, and optimize HPC code. The low-level features and software characteristics of even the most widely used libraries (e.g., math kernels, programming models) make the training and fine-tuning cumbersome to understand and reason with for \textit{science corner cases}. Heterogeneity adds more complexity owing to the multitude of vendors and runtime options that AI can be trained on. Additionally, because AI is statistical in nature, it adds its own bias, hallucinations, poor reproducibility, and errors. HPC software requires significant experimentation, and the data currently available from public sources (e.g., software repositories such as GitHub and GitLab, published literature) is not sufficient to train LLMs on the more important aspects beyond simple coding. 

\paragraph{Engineering the software ecosystem:} The DOE's ECP made significant investments that enabled the strategic application of software engineering practices \cite{10499282,10494039,GODOY2025107502}. These efforts included customized approaches such as continuous integration, performance testing on CPUs and GPUs, unified packaging and deployment via Spack~\cite{gamblin2015spack}, memory sanitizers, SLURM scripts, workflow composition, and more. Hence, the cost of these tasks must be commensurate with post-ECP funding levels. Current AI technologies are not trained on custom HPC solutions that target specific hardware and use cases for validation and verification~\cite{MUNLEY20241}. For example, we might have to adapt current testing knowledge to future architectures, and we might have to increase the coverage and capabilities of existing test programs and frameworks to match the evolution of the software. 

\paragraph{Ethics:} Several ethical challenges are associated with LLMs, including intellectual property and copyright, security and privacy, bias and fairness, and impact on software developer and skill degradation.

\paragraph{Training and workforce development:} State-of-the-art AI provides incredible opportunities in the broader landscape of science education~\cite{cooper2023examining}. The handful of core developers and advanced users are typically the \textit{educators} of HPC software. Like any specialized field, AI-generated educational resources need the critical oversight of those HPC software educators. Foregoing that oversight could actually increase the costs of training future HPC software developers because relying on unsupervised multimodality resources could lead to misuse of the software and a steeper learning curve in an already highly experimental field.

%% file: 04-Directions.tex
\section{Research Directions}
\label{sec:Research Directions}

Below, we outline future research directions for a more productive HPC software ecosystem to advance scientific discovery. These efforts are part of two ASCR AI4Science funded projects with two clear goals: 

\begin{enumerate}[(1)]
    \item Durban: AI for performance-portable programming systems and libraries
    \item Ellora: LLMs for a highly productive, AI-assisted HPC software ecosystem that targets E4S products
\end{enumerate}

These efforts aim to be cross-cutting and cover different aspects of the complex HPC software life cycle. As with the broader software landscape, state-of-the-art AI offers many opportunities for deployment at low (e.g., compilers, runtimes), intermediate (e.g., code generation, models), and high (e.g., software engineering, packaging, documentation) levels of the software development stack. 

\begin{figure}[htp]
\centering
%
\includegraphics[width=0.9\textwidth,height=0.5\textwidth]{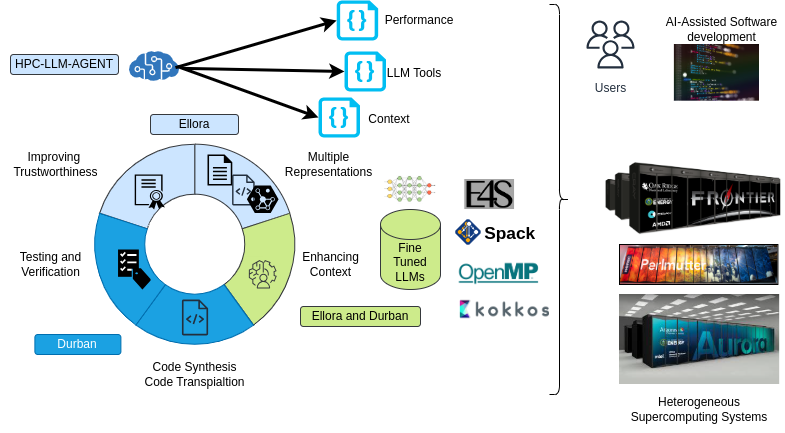}
\label{fig:ellora-durban}
\caption{Research thrusts for Durban and Ellora, which target ASCR's focus on AI Innovations for Scientific Knowledge Synthesis and Software Development.}
\label{fig:vision}
\end{figure}

We have identified several research directions to advance DOE's computing mission and to ensure that these innovations are mutually beneficial for the HPC and broader software development communities. Figure~\ref{fig:vision} shows the different research thrust areas that each project targets and their clear complementary efforts.

\paragraph{Fine-tuning LLMs:} 
To leverage state-of-the-art LLMs such as Llama and GPT, much effort is needed to introduce expert-in-the-loop knowledge. Specialized LLMs for coding such as CodeLlama and Deepseek-Coder~\cite{guo2024deepseek} demonstrate the potential for fine-tuning LLMs for a specific area.
Applying techniques from libraries such as LoRA (low-rank adaptation of LLMs)~\cite{hu2022lora} has shown on-par or better success in fine-tuned LLMs versus the 175-billion parameter GPT models. LoRA 
requires fewer resources (i.e., trainable parameters) while displaying a higher throughput and no additional inference latency. Preliminary work has also been done for the popular level-1 BLAS subroutines~\cite{10820659}, and this ChatBLAS effort showcases the path forward to integrating fine-tuning with HPC software. This direction requires targeted efforts to train foundational LLMs to understand patterns in HPC software APIs and contracts. Exploration of fine-tuning capabilities is a key research direction for an effective cost-benefit option.

\paragraph{Enhance context window:}
A context window, measured with ``tokens,'' is the limited text or memory that an LLM can process at a time. This limitation impacts HPC software largely because individual components are rarely used in isolation, and composing very specialized parallel simulation and analysis codes can be incredibly complex.
To overcome the constraints imposed by a model’s context window size, we must explore efficient solutions for large-context inference to reduce the likelihood of LLM hallucinations and increasing the likelihood of producing authoritative and relevant information. 

\paragraph{Leverage multimodality:}
Current coding LLMs are trained on code as text and thus lack information on program behavior (e.g.,
data flow, control flow) and are limited in associating additional metadata (e.g., performance characteristics) with the code structure, and this limits their ability to generate efficient HPC programs.
Being able to learn across multiple code modalities will improve the ability of coding LLMs to reason about code properties by giving them access to more data. We need to curate large-scale, multimodal code datasets that include code as text, LLVM Intermediate Representation (IR), performance counters etc. We also need to develop new, fine-tuned, multimodal LLMs that comprehend different code modalities to improve the performance on HPC coding tasks.
\paragraph{Enhance existing program source synthesis and compiler optimization methods, auto-tuning, and source-to-source transpilation:}
The primary goal of program source synthesis is to enhance the correctness of the generated source code based on the given prompt description. In other words, the generated code should align semantically with the prompt, thereby ensuring a strong correspondence between them. This alignment is often evaluated subjectively and requires some external processing for training to establish the context and most common patterns specific to the applications of interest. One good example is using a specific set of prompt keywords to navigate compiler optimization~\cite{italiano2025finding} incrementally to verify individual optimization techniques through differential testing. This approach can be extended to different programming languages and parallel programming frameworks to clarify their context and common patterns for training.   

Compared to prompt-based code synthesis, source-to-source transpilation involves less ambiguity. However, it has broad applications in the HPC and scientific computing communities, in which many scientific simulations still rely on large legacy code bases such as Fortran~77. Modernizing these code bases is essential for adapting to contemporary HPC architectures and system features, including massive parallelism, accelerators, asynchronous task execution, and the integration of modern HPC libraries for I/O and visualization. Additionally, an important goal of transpilation is optimizing application performance for modern HPC systems.  

For both modernization and performance optimization, LLM-based code transpilation must generate semantically equivalent programs while maintaining good code readability. Efforts to develop foundation models~\cite{tehranijamsaz_coderosetta_2024} for translating between HPC-relevant programming languages have successfully leveraged compiler abstract syntax trees to recognize patterns and apply language-specific keywords before the conventional training process. These preprocessing techniques can be extended to other source code analysis methods (e.g., semantic lifting~\cite{zhang2025semanticsliftingscientificcomputing} and lowering) to create training datasets at different levels of granularity. Formal methods would help ensure the semantic equivalence of two program sources~\cite{churchill19semantic}, but these techniques are still immature in parallel computing, and the scalability of formal checking needs special attention to make it feasible for large program sources.


\paragraph{Improve dynamic mapping, scheduling, and data orchestration:}
In heterogeneous computing architectures, performance and energy consumption often depend on the mapping of kernels to the most appropriate compute device(s) and on the orchestration of the data necessary for that kernel’s execution. To that end, AI could be used to explore a variety of techniques for auto-tuning, scheduling, and performance modeling. Reinforcement learning and graph neural network methods are excellent candidates for task and task-graph levels for code optimizations. Optimization of truly heterogeneous systems could be explored by leveraging AI and task-based HPC runtime frameworks (e.g., IRIS~\cite{10596490}).

\paragraph{Trustworthiness and verifiability:}
We must explore techniques that can characterize correct behaviors and also react to potential errors in the ever-evolving HPC software targets. Testing efforts that verify model correctness and ensure proper understanding of errors will reduce hallucinations and false positives in an AI-assisted software ecosystem.
Research directions should focus on improving prediction and explanation of model errors on coding tasks, including robust testing frameworks for verification and validation of AI-generated HPC software, and addressing other key challenges in this domain.

\paragraph{Integration and high-level interfaces:}
Parallel code development in HPC software is inherently an iterative process that
requires a developer to break down tasks into a sequence of steps and use various tools to complete
them. LLM agents have shown their potential in software development practices~\cite{talebirad2023multi}, and developing and using HPC-LLM agents could reduce the manual effort required to deliver an entire workflow if the LLM-based system can orchestrate
and execute complex, multistep tasks with the support of tools and developer feedback.

\paragraph{Ethics:} Ethical challenges associated with LLMs range from intellectual property and copyright to biases and fairness. Developing ethical guidelines for using LLMs is important if we want to ensure responsible use. Robust testing and validation are essential to ensuring the security and reliability of the generated code, and making LLMs more explainable can help identify potential biases. By considering all these aspects, we can work toward developing and using coding LLMs in a responsible fashion.

%% file: 05-Conclusions.tex
\section{Conclusions}
\label{sec:Conclusions}

Although today's unprecedented investments in AI have not yet inundated the HPC software ecosystem, which is characterized by low-data resources and high-specialization, 
we have provided our view for leveraging state-of-the-art AI technologies to advance current and future HPC software efforts.
We outlined the nuances of HPC software and the challenges and research directions needed to capitalize on the investments in AI.
These key research directions require multidisciplinary collaboration between HPC and AI experts in the larger software development community.
Although productivity is an important goal, any advancements must be accompanied with trustworthiness in the AI-assisted activities. Therefore, it is crucial that the community be involved in the research directions of a future AI-powered HPC software ecosystem.